# Retrosynthesis with Attention-Based NMT Model and Chemical Analysis of the "Wrong" Predictions

Hongliang Duan[§*], Ling Wang[§], Chengyun Zhang[§], Jianjun Li[*]

Artificial Intelligent Aided Drug Discovery Lab, College of Pharmaceutical Sciences, Zhejiang University of Technology, Hangzhou, 310014, PR China.

**ABSTRACT**

We cast retrosynthesis as a machine translation problem by introducing a special Tensor2Tensor, an entire attention-based and fully data-driven model. Given a data set comprising about 50,000 diverse reactions extracted from USPTO patents, the model significantly outperforms seq2seq model (34.7%) on a top-1 accuracy by achieving 54.1%. For yielding better results, parameters such as batch size and training time are thoroughly investigated to train the model. Additionally, we offer a novel insight into the causes of grammatically invalid SMILES, and conduct a test in which experienced chemists pick out and analyze the "wrong" predictions that may be chemically plausible but differ from the ground truth. Actually, the effectiveness of our model is underestimated and the "true" top-1 accuracy can reach to 64.6%.

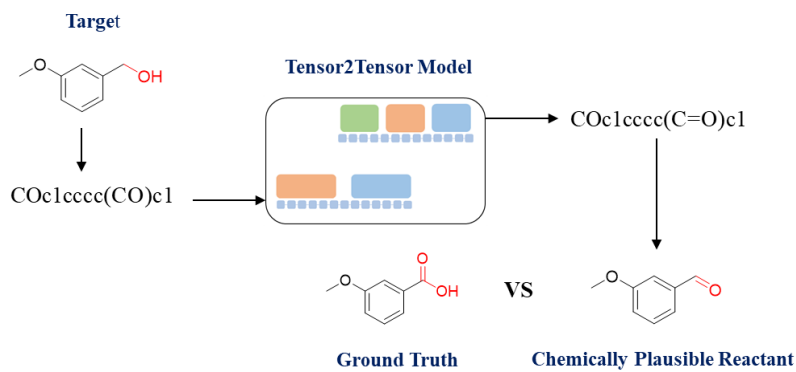

## INTRODUCTION

Organic synthesis is a crucial cornerstone of pharmaceutical, biomedical, and materials industries. There are two closely related issues about synthesizing new molecules: reaction prediction and retrosynthesis. In reaction prediction, the task is to deduce what could be the underlying product given a set of reaction building blocks such as reactants, reagents and reaction conditions. In retrosynthesis, it approaches this problem in reverse: starting from the target compound chemist hunger for creating, and exploring the simpler precursors commercially available[1-3] (Figure 1). By sequentially combining all the reactions derived from the retrosynthetic analysis, the overall synthetic route of the target molecule can be carried out.

Over the past few decades, multiple methods have been developed to carry out retrosynthetic analysis using novel and emerging computing techniques[4-6]. Computers have been used since the 1960s for storing chemical structures data, and for applying chemical structural information to the applications such as synthesis planning, drug discovery[7]. Corey and Wipke introduced computer assistance to chemical synthesis with Logic and Heuristics Applied to Synthetic Analysis (LHASA)[8] which is the first retrosynthesis program. They pioneered the use of expertly crafted rules regularly alluded to as reaction templates. Knowledge base-Oriented system for Synthesis Planning (KOSP) is another approach to empirical retrosynthesis. The system was built on the knowledge base in which reactions are abstracted by structural characteristics of reaction sites and their environments[9]. Generally, computer assisted retrosynthetic analysis is performed exploiting reaction rules that include a series of tiny transformations to characterize chemical reactions. These rules can either be laboriously encoded by chemical experts, or extracted from varieties of chemical digital data[10-18]. The outstanding advantage of the rules is that they can be interpreted directly. However, the rule-based methods remain several drawbacks. First, since there is currently no comprehensive rule system covering all chemical fields, rule-based systems cannot synthesize new compounds beyond their knowledge. Furthermore, the rules needed to be coded and curated, which is prohibitively expensive and timing consumed.

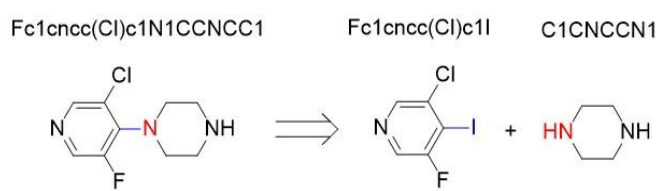

**Figure 1.** An example of retrosynthetic reaction: the target compound on the left of the arrow and the potential reactants on the right side are shown in common chemistry-like scheme and SMILES representation.

Deep learning (DL) is a class of machine learning algorithms that bring artificial neural network (ANN) containing multi-layer nonlinear processing unit to learn data representations[19]. Since the earliest ANN was established in 1943, significant improvements had been made during the 1960s to the 1980s[20]. Recent advances of DL in computer games and self-driving cars have demonstrated the potential applications throughout the world[21].

Given the increased availability of a great variety of digital data and algorithms, it's not surprising that we turn to DL for better use of reaction data on retrosynthesis. Recently, fully data driven approaches have been adopted to break through restrictions of rule-based systems. Molecule can be equivalently represented as a text sequence, such as the simplified molecular-input line-entry system (SMILES)[22]. From the perspective of linguistic, SMILES is regarded as a language. In this sense, the problem of chemical reaction can be treated as a translation task. Nam and Kim first introduced sequence-to-sequence (seq2seq) model, a neural machine translation (NMT) model, tested by examples and problems in an organic chemistry textbook of Wade to the reaction prediction. They mapped the SMILES representing the reactants to the SMILES representing the products[23]. Schwaller et al. built on the idea of relating reaction prediction to a language and explored the application of a neural machine translation method known as seq2seq model[24]. Retrosynthesis is the opposite of reaction prediction. Based on the premise that we approach the reaction prediction in reverse, there is a great possibility that seq2seq model is able to deal with retrosynthetic problem.

Liu et al. formulated retrosynthesis as a translation task using seq2seq model[25]. Given an input SMILES that represents the target compound, the model outputs a SMILES that represents the reactant. In this approach, the accuracy is 37.4% for top-1, 52.4 % for top-2 and 61.7% for top-5, which performed comparably or worse with a rule-based expert system baseline model. Meanwhile, the grammatical invalidity rate of the tpo-1 predicted SMILES is over 10%, Which limits the potential of application in future retrosynthetic reaction prediction.

Herein, we present an attention-based NMT model, Tensor2Tensor (T2T) model, which shows great superiority to machine translation task while being more parallelizable and requiring significantly less time to train[26]. Similar approaches were recently suggested[27-28]. In this paper, our team focus on the problem of retrosynthesis: "Given the target product, what are the most likely reactants?" The innovative T2T model is applied to retrosynthesis, which procure higher accuracy (54.1%) than previous work[25] and 3% invalidity of the top-1 predicted reactants SMILES on a common benchmark database. Diverse parameters such as batch size and training time are investigated to train the model, and we find that batch size should be set as high as possible while keeping a reserve for not hitting the out-of-memory errors and an even long training time can yield better performance. In addition, we deeply analyze incorrect SMILES and discover that two factors including the complexity of chemical structure and the lack of training data, may lead to the failure in the text presentation. We also conduct a test in which 10 experienced chemists pick out and analyze the "wrong" predictions that may be true from chemists' point of view but not match the ground truth.

**APPROACH**

**Data.** We adopt the same dataset of reactions as in Liu's work. The dataset containing reaction examples derived from USPTO patents was originally prepared by Lowe[29]. Schneider et al. extracted 50,000 reactions spanning 10 broad reaction types (Table S1) from it to represent the typical reaction types found in the medical chemist's toolkit and discarded contextual information (e.g., temperature, reagents, yields) for only comprising reactants and products[30]. Additionally, Liu et al. further developed this dataset by splitting multiple product reactions so that each reaction example contains a single product. Finally, there are 40029 reactions for training, 5004 reactions for validation, 5004 reactions for testing and all reactions are described by a text-based representation called SMILES.

**Model.** T2T model is implemented to retrosynthetic reaction prediction. This model based on an encoder-decoder architecture initially was constructed for neural machine translation tasks[31] and it shows state of the art performance in chemical reaction prediction[32].

The architectural characteristic of T2T model is that it entirely depends on attention mechanisms. As a new generation of encoder-decoder neural network model, T2T model, comprising feed-forward network and multi self-attention layers, avoids complicated recurrent or convolutional neural networks. It can get queries ($Q$) that data inquire, search keys ($K$) for the indexed knowledge and acquire values ($V$) related to queries and keys, then matrixes learn them during training. In order to obtain queries ($q$), keys ($k$), values ($v$) corresponding to a current batch, the T2T model multiply $Q$, $K$, $V$ with the input ($X$). With these computed parameters, the inputs can be transformed to some encoding parts or decoding parts.

As depicted in Figure 2, the main components of the model are encoder and decoder stacks. The encoder is composed of several same layers and each layer contains two different sub-layers. The first sub-layer is a multi-head self-attention mechanism and the second is a feed-forward network layer. Before layer normalization[33], a residual connection[34] is applied around each of the two sub-layers. The decoder consists of identical layers but each layer is comprised of three different sub-layers. Apart from the two sub-layers mentioned, there is a third sub-layer called masked multi-head self-attention mechanism, and the residual connection still is employed around each of the sub-layers as well as the encoder.

Remarkably, the multi-head attention which consists of parallel attention layers is an innovative part of T2T model. We can perform the attention function in parallel to get different versions of output values after linearly projecting the queries, key, and values, then concatenate and again project them to obtain the final values. Hence, the model with several sets of independent attention parameters outperforms models with a single attention function.

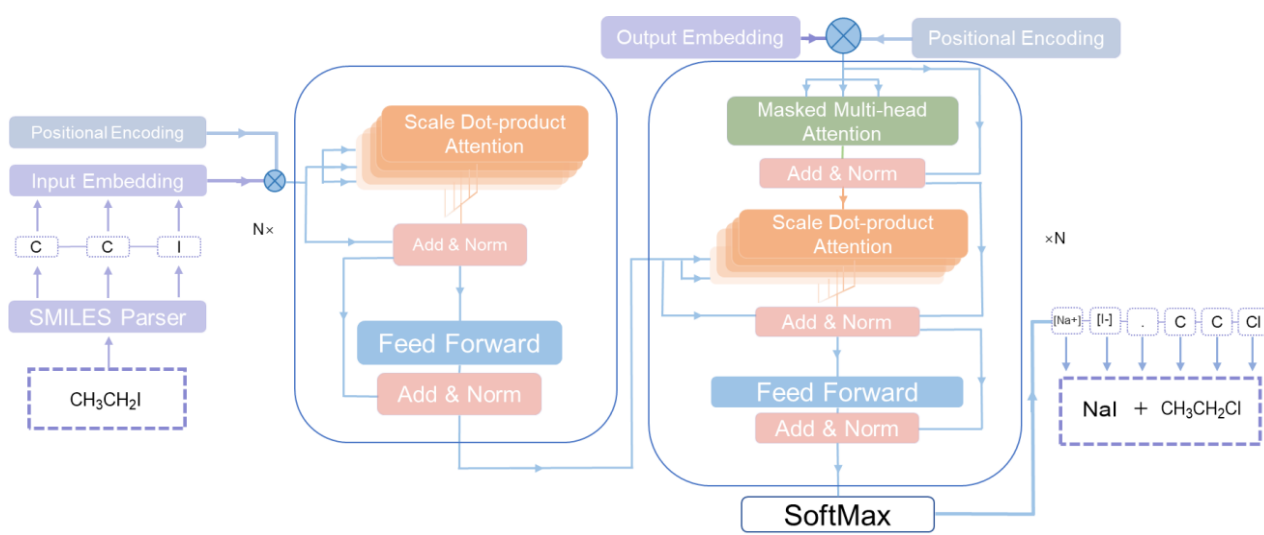

**Figure 2.** T2T model architecture. Inputs and outputs are described by SMILES in this model. The encoder(left) and the decoder(right) consist of a stack of N identical layers. N is the number of layers.

Several scaled dot-product attention layers make up a multi-head attention layer. They take the input made up of queries, keys of dimension *dk*, and values of dimension *dk,* then calculate the three entities. The dot products of the query are computed with all keys to explore alignment between the keys with them, then we multiply the results by $\frac{1}{\sqrt{dk}}$, and get the weights of the values. The queries, keys, and values that the attention compute will be packed together into matrices in practice so that we can get a matrix of outputs.

T2T model pays same attention to the elements of a sequence no matter how long distances between tokens, resulting in information about the relative or absolute position of tokens in sequences may be missing. A positional encoding matrix was proposed to solve such problems. Depending on the position in a sequence and in the embedding direction, the elements equal to the values of trigonometric functions. The positional encoding make connection between far located parts of the input with learned embedding. With it, the model can make use of the order of the sequence.

**Baseline model**. Baseline model is seq2seq model[35] which was employed by Liu et al. to perform retrosynthetic reaction prediction. The primary parts of the model are the encoder and decoder part which are both made up of long short-term memory cells[36], a variant of recurrent neural network cells. In the architecture, the encoder takes sequences of the input and trains it then passes corresponding context vector to the decoder, and the decoder uses the representation and gives sequences of the output.

However, there are some problems in this model. One major drawback of seq2seq model is that with recurrence operation, the computation can't be parallelized on multi GPUs. In addition, the main limitation in the architecture's ability to process sequences is the size of information the fixed-length encoded feature vector can contain. And the performance of the model decreases as the length of sequences increases. Therefore, it's rather challenging for seq2seq model to tackle too long sequences.

All program scripts were implemented in Python (version 3.5) and the open-source cheminformatics toolkit RDKit[37] (version 2019.03.10) was applied for the deeply analyze the invalidity of SMILES. Our model was built with TensorFlow [38](version 1.11.0)

## RESULT

**Comparison of accuracies between the baseline model and T2T model.** The prediction accuracies of our model and baseline model on Top-N are described in Figure 3. Our model is superior to baseline model by a margin of 13.8% on top-1 accuracy. With the increase of N, there is a prominent improvement in accuracies of those models: the top-10 accuracy of T2T model achieves 70.1% and baseline model earns 61.7%.

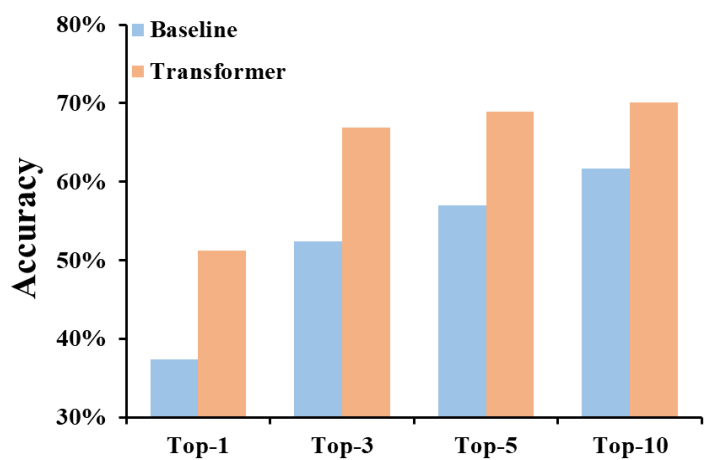

**Figure 3.** Top-N accuracies of T2T model and baseline model on test data set.

In addition, the performance of our model in different reaction types is also explored. We compare the results to the baseline model and in all of the reaction types our model performs significantly better. Take reaction class 3 (C-C bond formation) as an example (Figure 4), the accuracy of C-C bond formation reactions in our model achieves 58.0%, which is 11.9% higher than the result of baseline model (46.1%). It is worth noting that reaction class 4, which contains the formation of cyclic structure (Figure 5), usually leads to a great variation between the reactants and products. Even for an experienced chemist, it is a tough retrosynthetic problem to decide the proper disconnection bond for the ring. While the result is not as good as in other reaction types, mainly due to the complexity of cyclic structure, our model achieves an accuracy of 54.4% surpassing noticeably baseline model.

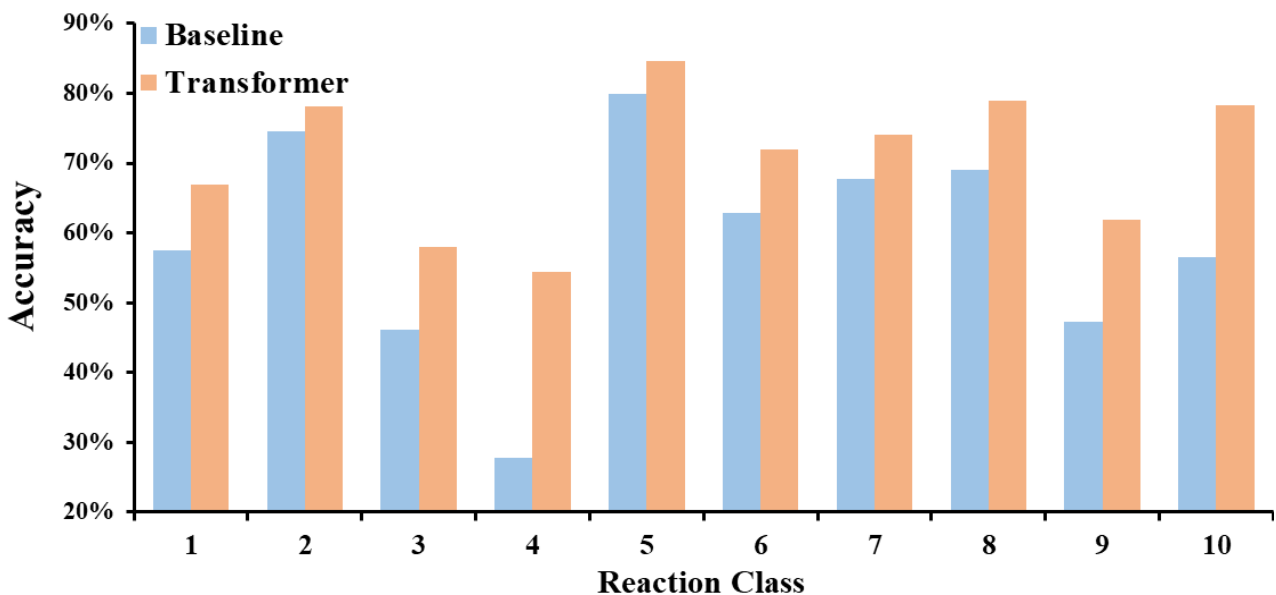

**Figure 4.** Top-10 accuracies of T2T and baseline model in different reaction classes.

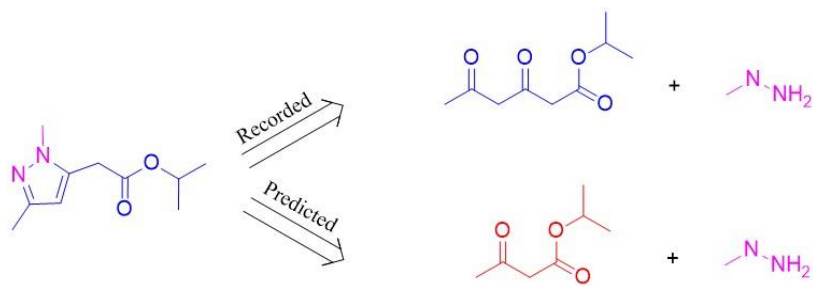

**Figure 5.** A representative example of prediction for cycle formation reaction.

**The effect of batch size.** Batch size is the number of training samples processed in one training step[39]. In our model, the parameter *batch_size* is defined as a representation of the amounts of tokens in a batch. As a crucial parameter, batch size affects the prediction quality, training speed and training stability in other neural network architectures[40]. To evaluate the influence of batch size on our attention-based model, we test our hypothesis by doing a set of experiments with varying *batch_size* within range 512-8192.

As depicted in Figure 6, after 10 hours of training, we get an accuracy of 43.4% with *batch_size* of 512 and 49.2% with *batch_size* of 2048, indicating that bigger batch sizes perform better than smaller ones. However, the accuracy does not significantly increase any more when the batch size exceeds 2048. For example, after 10 hours of training, an accuracy of 50.1% with *batch_size* of 4096 is achieved compared with 50.6% with *batch_size* of 6144. There is no substantial difference of accuracies between batch size 4096 and 6144.

The explanation of this phenomenon is that training throughput, which is the number of training data processed in the training, affects markedly the performance of our model with a high batch size. To our knowledge, the bigger the batch size is, the slower computation speed is, and as batch size become bigger, training throughput which equals to multiply batch size by computation speed improves slightly[39]. Therefore, the predictive capabilities don't increase greatly when batch size exceeds a certain size, as a result of mildly higher throughput. Note that setting batch size too big may results in the out-of-memory errors and setting it too lower leads to notably low accuracy. Thus, it's of a great advantage to employ proper batch size.

**The influence of training time.** In our experiments, the accuracy commonly doesn't grow greatly any more after several hours of training but it's not clear whether an even longer training time can lead to better results. To explore the effect of training time on our model, a test with a training time as long as five days on our GPU is carried out.

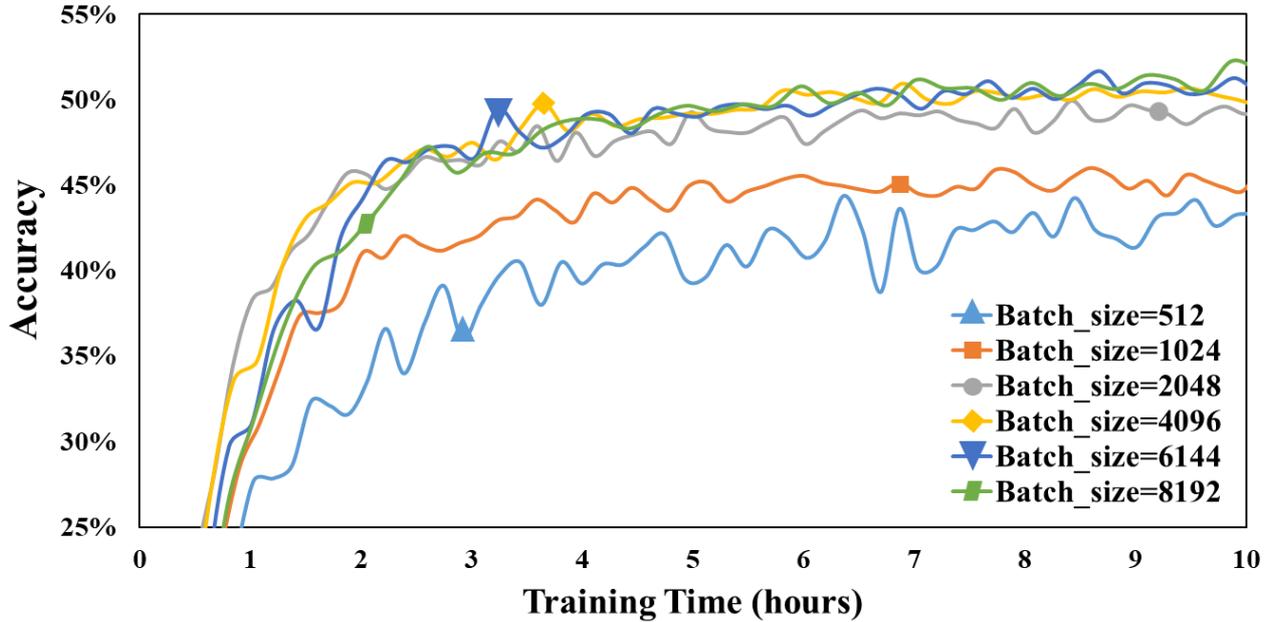

**Figure 6.** The accuracies of T2T model with different batch sizes. Note that all is trained and tested on a 1080Ti GPU.

**Table 1. Influence of Batch Size and Training Time on Performance.**

| Batch size | Training time | Accuracy (%) |
|---|---|---|
| 512 | 10 h | 43.4 |
| 1024 | 10 h | 45.7 |
| 2048 | 10 h | 49.2 |
| 4096 | 10 h | 50.1 |
| 6144 | 10 h | 50.6 |
| 8192 | 10 h | 51.8 |
| 8192 | 5 d | 53.0 |
| 8196.Avg | 5 d | 54.1 |

As illustrated in Table 1, with *batch_size* of 8192, the accuracy can reach to 51.8% after 10 hours of training, and the accuracy is further improved to 53.0% after 5 days training. It shows that the model earns a marginally better result after long time training. We conduct an additional experiment by averaging checkpoint to improve accuracy. By averaging the last 20 checkpoints saved in 2000-steps intervals and training the model for 5 days, we achieve the accuracy of 54.1% with *batch_size* of 8192. Compared to previous work, the architecture employed checkpoint averaging can predict the correct answers with higher probability, meaning that it's advisable to apply checkpoint averaging to our model.

**DISCUSSION**

**Grammatically invalid SMILES comparison between the baseline model and T2T model.** Invalidity rate of the top-1 predicted reactants SMILES strings in T2T model is 3.4%, yielding better results than seq2seq model (12.2%). Liu et al. attached great importance to reaction types in consideration what factors cause the seq2seq model to make a lot of grammatical mistakes in the SMILES predictions[25]. In the course of our study, there is an intriguing contradiction with Liu' view that the validity rate of SMILES is comparable in each reaction class (Figure S1), namely, reaction types are not related to the invalidation of SMILES.

We profoundly analyze all compounds which are predicted to grammatically invalid SMILES by seq2seq model and T2T model. There are two main factors including the complexity of chemical structure and the lack of training data, may cause models to predict text representation incorrectly.

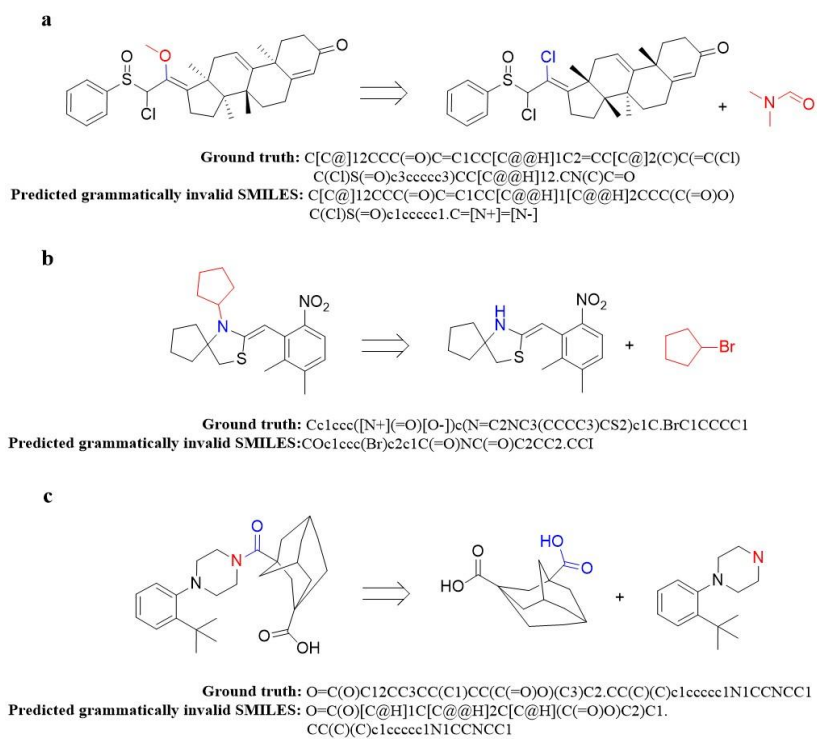

**Figure 7.** Examples of three typical types of cyclic compounds which are prone to be predicted to grammatically invalid reactants SMILES. (a) Polycyclic aromatic hydrocarbons. (b) Spirocyclic hydrocarbons. (c) Bridged hydrocarbons.

Especially for complicated cyclic compounds including polycyclic, spirocyclic and bridged hydrocarbons, the models generally output invalid SMILES during the retrosynthetic analysis (Figure 7). The key feature of these cyclic compounds is perplexing ring structure units, and it is a challenging assignment for excellent chemists to name them. Taking spirocyclic hydrocarbon as an example, the systematic naming of it based on IUPAC rules is rather tough due to its excessive complexity. Moreover, the lack of relative reaction examples also leads to wrong reactants SMILES.

In addition, quaternary carbon structures could have tremendous influence on performance of seq2seq model. The task, predicting valid reactants SMILES about molecules containing Boc, $CF_3$, $C(CH_3)_3$, is tough for seq2seq model but is a trivial problem for T2T model. As shown in Table 2, there are 71 predictions containing the structures mentioned above and account for 11.6 of the total invalid SMILES strings in seq2seq model. But our model basically does not go wrong on the prediction of this class compounds. Figure 8 shows representative examples which T2T model is capable of predicting SMILES correctly but seq2seq model fails.

**Table 2. The Distribution of Predictions[a] about Molecules Containing Quaternary Carbon Structure.**

| Structure | Count | Rate (%) |
|---|---|---|
| R-C(=O)-O-C(CH3)3 (Boc) | 38 | 6.2 |
| R-CF3 | 18 | 2.9 |
| R-C(CH3)3 | 15 | 2.5 |
| **Total** | **71** | **11.6** |

[a] T2T model output SMILES correctly while seq2seq model is unable to predict correct answers.

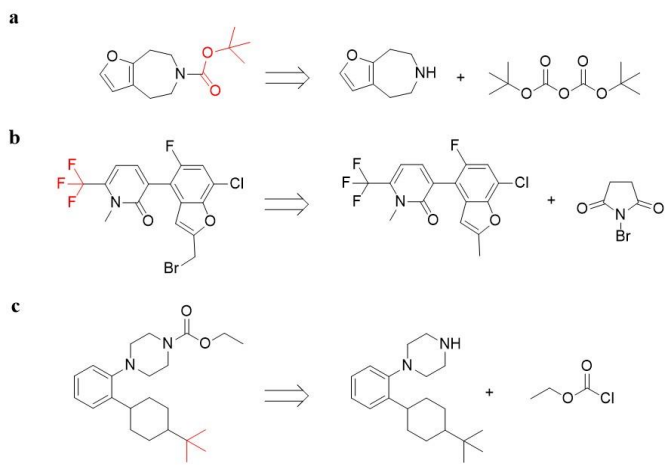

**Figure 8.** Characteristic examples which T2T model is able to predict valid reactant SMILES while seq2seq model fails. (a) Compound containing Boc. (b) Compound containing $CF_3$. (c) Compound containing $C(CH_3)_3$.

**Chemical analysis of the "wrong" predictions.** The accuracy of model refers to the proportion of ground truth reactants is found in the predictions predicted by the model. In our work, we are aware of limitation of the accuracy metric: a lot of chemically plausible answers which have been proven to be feasible in the literature or have not been discovered, are judged as "wrong" answers, which result in lower accuracy of the model. To quantitatively appraise this, we conduct a test which contains 10 chemists with rich experience in organic chemistry. To pick out the "wrong" answer, the participants resort to repositories such as SciFinder[41], Reaxys[42], which are web-based tools for the retrieval of chemistry information and data from published literature. Representative examples of chemically plausible predictions are shown in Figure 9.

The simplest method of synthesizing acid is hydrolysis of carboxylic ester. For example, both benzoic acid methyl ester and benzoic acid phenylmethyl ester can form benzoic acid by hydrolysis[43]. Noted in Figure 9a, the recorded outcome shows that the target compound could be synthesized by hydrolysis of corresponding methyl ester. The prediction is consistent with the method of Leggio et al.[44] (Figure 10a) which employed hydrolysis of corresponding benzyl ester to produce the target compound.

Oxidation is a large class of chemical reaction in which the atoms of an element lose electron. Figure 9b illustrates retrosynthetic analysis of 2-(trifluoromethyl) pyridine which is the key intermediate of B Raf inhibitory. There are a number of oxidants containing electro-philic oxygen atoms that react with nucleophilic pyridine to produce the target compound. The most commonly used oxidation agents are meta-chloroperoxybenzoic acid[45] (mCPBA) and $H_2O_2$ which are actually comparable to each other in this reaction. The prediction is chemically plausible due to the oxidizing agent, mCPBA, but that is likely missed in our recorded reaction examples. As a matter of fact, Aquila [46] have reported an approach (Figure 10b) to the target compound where the mCPBA was employed, and has made an outstanding contribution to the synthesis for B Raf inhibitory.

During the preparation of complex organic molecules, there are often stages in which one or more protecting groups are introduced into the molecule in order to protect the functional groups from reacting, and won't be removed until the reaction is completed. The first stage of this reaction is protection, and the second stage is deprotection. Synthesizing the target molecule is a stumbling block on the way to synthesize the intermediate of TORC1/2 inhibitor. Figure 9c describes a reaction example where the precursor is deprotected to form the target compound. In addition to the benzyl group used in the recorded outcome, other alternative protecting groups, such as Boc, Cbz, Fmoc, etc, could be applied to protect amino from powerful electrophile. The prediction adopting the Boc protecting group is not captured in our recorded example, which is mistakenly considered to be wrong answer, but it has been confirmed to be chemically plausible by Hicks et al.[47] (Figure 10c).

Metal-catalyzed cross-coupling reactions including Stille Coupling, Suzuki Coupling, etc, are common reactions to form C-C bond, which play a significant role in organic synthesis. For instance, the recorded outcome shows that the target compound could be formed by Stille Coupling between trimethyl (4-methylphenyl) stannane and 1-acetyl-2-bromobenzene (Figure 9d). The Stille Coupling is a versatile C-C bond forming reaction between stannanes and halides, with very few limitations on the R-groups. While the predicted outcome displays that the target compound could be synthesized by Suzuki reaction between 1-bromo-4-methylbenzene and (2-acetylphenyl) boronic acid. This was also the route that was chosen by Laha et al.[48] (Figure 10d). The above two methods which are substitutive for each other on a large scale have profoundly changed the protocols for the construction of chemical molecules.

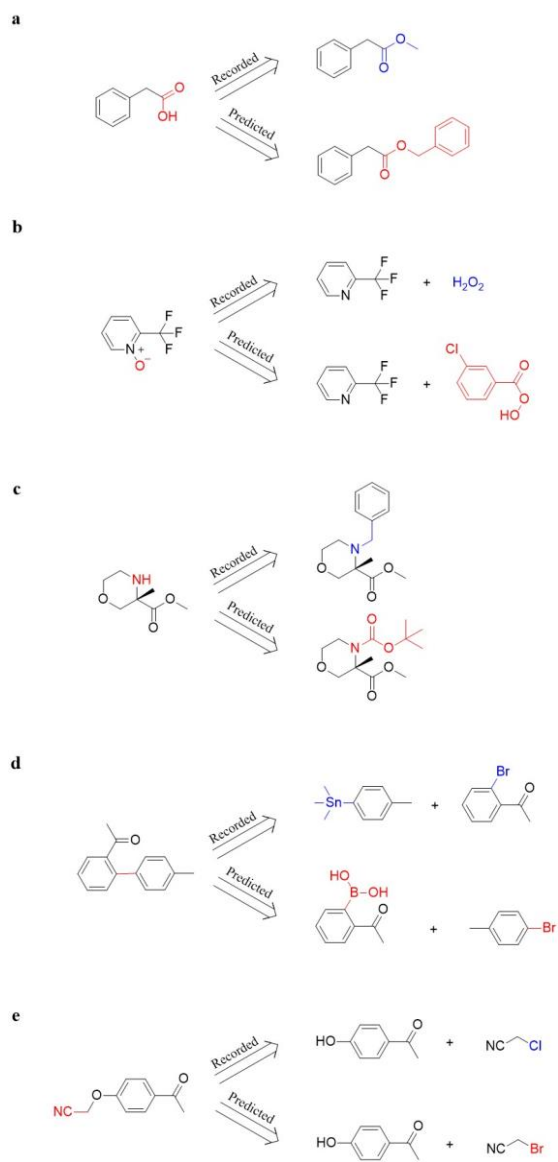 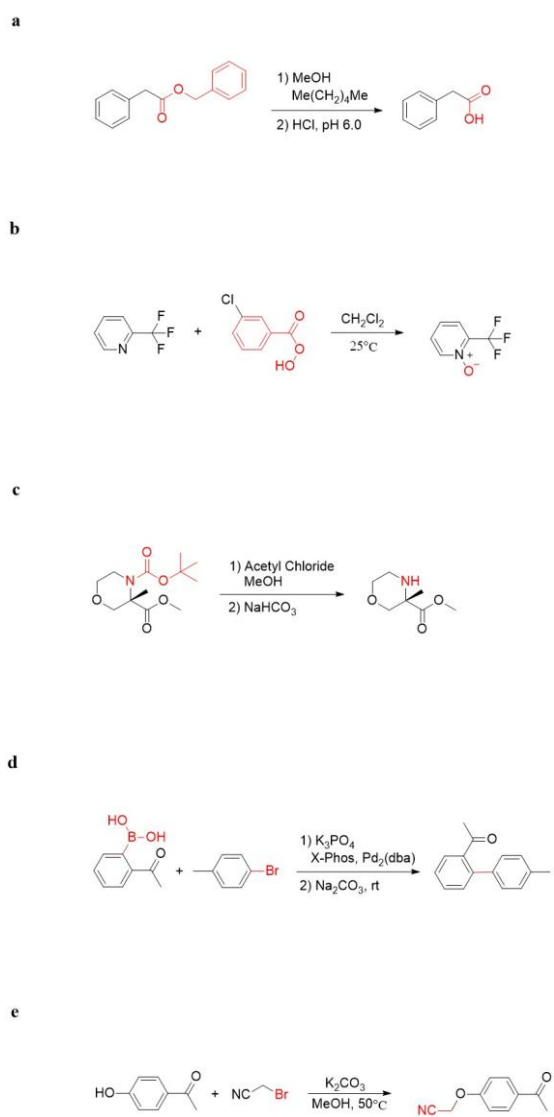

**Figure 9.** Examples of chemically plausible predictions considered as "wrong" answers. (a) Hydrolysis from varied carboxylic ester to acid. (b) Oxidation by different oxidant. (c) Protection with diverse protecting group. (d) C-C bond formation via various cross-coupling reaction. (e) $S_N2$ reaction between alkoxide (amine) and alkyl halide.

**Figure 10.** Examples of "wrong" predicted reactions matching the published routes. (a) Hydrolysis from benzyl ester. (b) Oxidation by mCPBA. (c) Protection with Boc. (d) C-C bond formation via Suzuki reaction. (e) A $S_N2$ etherification reaction between alkoxide and bromoalkene.

A universal way to make ethers is to treat alkoxide anions with halohydrocarbons including chloroalkane, bromoalkene and iodoalkane. It is a remarkable fact that the target molecule is the pivotal intermediate of bazedoxifene acetate used as drug for treatment or prevention of postmenopausal women osteoporosis. As the two halohydrocarbons can be replaced mutually in $S_N2$ reaction, there is no intrinsic difference between the recorded reaction and predicted reaction. The ground truth shows a simple $S_N2$ etherification reactions where 1-(4-hydroxyphenyl) ethenone react with 2-chloroacetonitrile to form the target compound and the prediction also displays a similar reaction which matches the published route[49](Figure 9e and Figure 10e). In this work, we provide an important method of higher yield and easier post-processing for the synthesis of key intermediates of bazedoxifene acetate.

Due to space constraints, we will not describe all chemically plausible reaction types in detail. Statistical measures are more intuitive than explaining the reaction examples separately. The number of various types of reactions is shown in Table 3. It's worth noting that examples of hydrolysis from varied carboxylic ester to acid are the largest of the seven types of reactions. There are two additional types of reactions, condensation between carboxylic acid (acid halide) and amine and reductions from disparate carbonyl compound to alcohol, that warrant mention, with high percentage 2.3% and 1.4%.

**Table 3. Breakdown of the Chemically Plausible Predictions for Different Reaction Types.**

| Chemically plausible reaction type | Count | Rate (%) |
|---|---|---|
| Oxidation by different oxidant | 11 | 0.2 |
| Protection with diverse protecting group | 50 | 1.0 |
| Hydrolysis from varied carboxylic ester to acid | 127 | 2.5 |
| C-C bond formation via various cross-coupling reaction | 36 | 0.7 |
| $S_N2$ reaction between alkoxide (amine) and alkyl halide | 123 | 2.4 |
| Reductions from disparate carbonyl compound to alcohol | 69 | 1.4 |
| Condensation between carboxylic acid (acid halide) and amine | 120 | 2.3 |
| **Total** | 536 | 10.5 |

**CONCLUSION**

In this work, we present a fully data-driven, solely attention-based T2T model, for retrosynthesis. The model outperforms the seq2seq model by a large margin in every class and has essential advantages, over conventional rule-based systems and any deep learning approach. This indicates that our approach clearly enhances the performance of DL for retrosynthetic reaction prediction task relative to all-rules machine learning models. However, a critical limitation about this accuracy metric is that a lot of chemically plausible prediction are judged as "wrong" answers, which results in a lower accuracy compared with the "true" accuracy. In our approach, there are 536 "wrong" predictions which account for 10.5% of the total test data set. Actually, the "true" accuracy of the model can reach to 64.6%. The inability of the accuracy metric to match our model highlights an important limitation of all current models. One way to alleviate this issue should be put on the agenda of researchers.

There are clear opportunities for application of T2T model to reaction prediction and retrosynthesis. While few researchers currently work in this area, we anticipate a dramatical increase in the coming years, as practical challenges of data availability are addressed.

## AUTHOR INFORMATION

**Corresponding Author**
*__E-mail:__ hduan@zjut.edu.cn
*__E-mail:__ lijianjun@zjut.edu.cn

**Author Contributions**

§Hongliang Duan, Ling Wang and Chengyun Zhang contributed equally.

**Notes**

The authors declare no competing financial interest.

## ACKNOWLEDGMENTS

We acknowledge the generous support of chemists from College of Pharmaceutical Sciences, Zhejiang University of Technology for our work on chemical analysis.